\documentclass[preprint,aps,nofootinbib]{revtex4}

\newcommand {\bc}{\begin{center}}
\newcommand {\ec}{\end{center}}
\newcommand {\bea}{\begin{eqnarray}}
\newcommand {\eea}{\end{eqnarray}}
\newcommand {\be}{\begin{equation}}
\newcommand {\ee}{\end{equation}}

\def\lsim{\mathrel{\rlap{\lower4pt\hbox{$\sim$}}
    \raise1pt\hbox{$<$}}}               
\def\gsim{\mathrel{\rlap{\lower4pt\hbox{$\sim$}}
    \raise1pt\hbox{$>$}}}  
\usepackage{graphicx}

\begin{document}


\title{Hydrodynamic fluctuations and the minimum shear viscosity 
of the dilute Fermi gas at unitarity}

\author{Clifford Chafin and Thomas~Sch\"afer}

\affiliation{Department of Physics, North Carolina State University,
Raleigh, NC 27695}

\begin{abstract}
We study hydrodynamic fluctuations in a non-relativistic fluid. 
We show that in three dimensions fluctuations lead to a minimum 
in the shear viscosity to entropy density ratio $\eta/s$ as a function of 
the temperature. The minimum provides a bound on $\eta/s$ which is 
independent of the conjectured bound in string theory, $\eta/s
\geq \hbar/(4\pi k_B)$, where $s$ is the entropy density. For
the dilute Fermi gas at unitarity we find $\eta/s\gsim 0.2\hbar$.
This bound is not universal -- it depends on thermodynamic 
properties of the unitary Fermi gas, and on empirical information
about the range of validity of hydrodynamics.  
We also find that the viscous relaxation time of a hydrodynamic
mode with frequency $\omega$ diverges as $1/\sqrt{\omega}$, and 
that the shear viscosity in two dimensions diverges as $\log(1/
\omega)$.       

\end{abstract}

\maketitle

\section{Introduction}
\label{sec_intro}

 It is now widely appreciated that fluid dynamics can be viewed
as an effective long-distance theory for a classical or quantum
many-body system at non-zero temperature. Effective theories 
make systematic predictions for correlation functions order-by-order
in a low-momentum expansion. These predictions depend on a small 
number of microscopic parameters. In the case of fluid dynamics
the microscopic parameters are the equation of state and the 
transport coefficients. 
 
 Effective (field) theories are constructed using the following 
procedure: i) Identify the low energy degrees of freedoms. ii) Write 
down the most general local effective action consistent with the 
symmetries of the problem. This action is typically expressed in 
terms of low energy fields and their derivatives. The coefficients
of allowed terms in the effective action are free parameters called 
low energy constants. iii) Determine what terms in the effective 
action have to be included in order to compute a correlation function 
to a given order in the low energy expansion. This is known as the 
``power counting''. Typically, the leading contribution arises from 
tree level diagrams involving operators with the minimal number of 
derivatives, and higher order corrections arise both from higher 
derivative operators, and from loop diagrams generated by the leading 
order interactions. In some cases diagrams may have to be summed to 
all orders. For example, the sum of all tree diagrams corresponds to 
solving a classical field equation. 

 In fluid dynamics the low energy modes are fluctuations of 
the conserved charges. In the case of a one-component non-relativistic  
fluid the conserved charges are the particle density, the energy density, 
and the momentum density. The derivative expansion is implemented at
the level of the constitutive equations. This means that the conserved
currents are expressed in terms of derivatives of the thermodynamic
variables. The effective theory without derivative terms in the 
currents is called ideal fluid dynamics, and the equation of motion 
at one-derivative order is known as the Navier-Stokes equation. The 
validity of fluid dynamics requires that derivative corrections to 
the currents are small. This condition does not preclude the possibility
that small corrections can exponentiate as one solves the equations of 
motion. Solutions of the Navier-Stokes equation are qualitatively 
different from solutions of ideal fluid dynamics. In ideal fluid 
dynamics the motion is time reversible, sound modes are not damped, 
and diffusive modes do not exist. This implies that in most cases one 
has to retain at least one-derivative terms in the constitutive equations. 
Two derivative terms have also been studied \cite{Garcia:2008,Chao:2011cy}, 
but the corrections are typically small. In relativistic fluid dynamics 
two-derivative terms improve the stability of equations of motion, and 
second order terms are now routinely included in hydrodynamic simulations
of relativistic heavy ion collisions \cite{Romatschke:2009im}.

 Loop corrections in fluid dynamics arise from thermal fluctuations.
Fluctuations are known to be important in the vicinity of second
order phase transitions \cite{Hohenberg:1977ym}, but they are rarely 
considered in the case of non-critical fluids. In this work we will 
consider the contribution of fluctuations to the correlation function 
of the stress tensor in a simple non-relativistic fluid. We study 
the implications of our results for the shear viscosity of the unitary 
Fermi gas. The unitary Fermi is known to have a very low viscosity 
\cite{Schafer:2007pr,Cao:2010wa,Schafer:2009dj,Adams:2012th}, close to 
the value obtained from the AdS/CFT (Anti-deSitter Space/Conformal 
Field Theory) correspondence, $\eta/s=\hbar/(4\pi k_B)$ 
\cite{Policastro:2001yc,Kovtun:2004de}. Here, $\eta$ is the shear 
viscosity and $s$ is the entropy density, $\hbar$ is Planck's constant
and $k_B$ is Boltzmann's constant. We will set $\hbar=k_B=1$ in the 
following. We will show that the small shear viscosity enhances the 
role of fluctuations, but we also show that fluctuations imply a lower 
limit on how small the viscosity can get. We demonstrate that fluctuations 
lead to a non-analytic term in the viscous relaxation time in three 
spatial dimensions, and to a logarithmic divergence of the shear 
viscosity in two dimensions. Finally, we discuss the possibility 
of observing these non-analytic terms in experiments with trapped
atomic gases. 

 Our work builds on a substantial literature related to fluctuations
in fluid dynamics, beginning with the work of Landau \cite{Landau:smII}.
The role of fluctuations in critical transport phenomena was summarized
in the review article by Hohenberg and Halperin \cite{Hohenberg:1977ym}
and the text books by Ma and Onuki \cite{Ma:1976,Onuki:2002}. Diagrammatic
methods are discussed by a number of authors, for example in 
\cite{Martin:1973zz,DeDominicis:1977fw,Khalatnikov:1983ak}. Our work
closely follows a recent study of fluctuations in relativistic fluids,
see \cite{Kovtun:2011np,PeraltaRamos:2011es}, the recent review 
\cite{Kovtun:2012rj}, and the related work in \cite{Torrieri:2011ne}.

\section{Kubo Formula}
\label{sec_kubo}

 In this section we will determine the low energy behavior of the 
retarded correlation function of the stress tensor using the classical 
equations of fluid dynamics at next-to-leading order in the gradient 
expansion. This result can be used to derive the standard Kubo formula for 
the shear viscosity, as well as a new Kubo formula for the viscous relaxation 
time. We will employ the formalism developed in 
\cite{Son:2005rv,Son:2005tj,Chao:2010tk,Chao:2011cy}, which is based 
on coupling the theory to a non-trivial background metric $g_{ij}(t,\vec{x})$. 
Correlation functions of the stress tensor can be computed by using linear 
response theory, and the requirements of Gallilean and conformal symmetry 
can be incorporated by requiring the equations of fluid dynamics to 
satisfy diffeomorphism and conformal invariance. 

 The retarded correlation function of the stress tensor $\Pi^{ij}$ is 
defined by
\be
\label{G_R}
G_R^{ijkl}(\omega,{\bf k}) = 
     -i \int dt \int d{\bf x}\, e^{i\omega t - i{\bf k \cdot x}} 
   \Theta(t) \langle [\Pi^{ij}(t,{\bf x}), \Pi^{kl}(0,{\bf 0})]\rangle \, . 
\ee
$G_R$ determines the stresses induced by a small perturbation $g_{ij}(t,
{\bf x})=\delta_{ij}+h_{ij}(t,{\bf x})$ around the flat metric. We have
\be
\label{G_R_lin}
\delta \Pi^{ij} = 
    - \frac{1}{2} G_R^{ijkl}h_{kl} \, . 
\ee
In fluid dynamics we expand the stress tensor in derivatives of
the local thermodynamic variables $P,\rho,{\bf v}$, where $P$ is the 
pressure, $\rho$ is the density, and ${\bf v}$ is the fluid velocity. 
We write $\Pi_{ij} = \Pi_{ij}^0+\delta\Pi_{ij}$, where
\be 
\Pi_{ij}^0 = \rho v_i v_j +P g_{ij} 
\ee
is the ideal fluid part, and $\delta\Pi_{ij}$ is the viscous correction. 
In a conformally invariant fluid the leading term is  $\delta\Pi_{ij}=
-\eta\sigma_{ij}$ with
\bea
\label{def_sig}
 \sigma_{ij} &=& \nabla_{i}v_{j}+\nabla_{j}v_{i}+\dot{g}_{ij}
         -\frac{2}{3}g_{ij}\langle\sigma\rangle \, , \\
\label{def_s}
 \langle\sigma\rangle &=& \nabla\cdot v+\frac{\dot{g}}{2g}\, , 
\eea
where $\sigma_{ij}$ is the shear stress tensor, $\eta$ is the shear 
viscosity, $g_{ij}\langle\sigma\rangle$ is the bulk stress tensor, and 
$\nabla_i$ is the covariant derivative associated with $g_{ij}$. Note 
that the bulk viscosity of a conformal fluid is zero. In \cite{Chao:2011cy} 
we classified all terms up to second order in derivatives. We have 
\bea 
\delta\Pi_{ij} &=& -\eta\sigma_{ij}
   + \eta\tau_R\left(
    g_{ik}\dot\sigma^{k}_{\; j} + v^k\nabla_k \sigma_{ij}
    + \frac{2}{3} \langle \sigma\rangle \sigma_{ij} \right) 
    + \lambda_1 \sigma_{\langle i}^{\;\;\; k}\sigma^{}_{j\rangle k} 
    + \lambda_2 \sigma_{\langle i}^{\;\;\; k}\Omega^{}_{j\rangle k}
   \nonumber \\
   && \mbox{} 
    + \lambda_3 \Omega_{\langle i}^{\;\;\; k}\Omega^{}_{j\rangle k}  
    + \gamma_1 \nabla_{\langle i}T\nabla_{j\rangle}T
    + \gamma_2 \nabla_{\langle i}P\nabla_{j\rangle}P
    + \gamma_3 \nabla_{\langle i}T\nabla_{j\rangle}P  \nonumber \\
   \label{del_pi_fin}
   && \mbox{}
    + \gamma_4 \nabla_{\langle i}\nabla_{j\rangle}T 
    + \gamma_5 \nabla_{\langle i}\nabla_{j\rangle}P
    + \kappa_R  R_{\langle ij\rangle}\, , 
\eea
where $\tau_R$ is the viscous relaxation time, $\lambda_{i}$, $\gamma_{i}$, 
$\kappa_R$ are second order transport coefficients, $\Omega_{ij}=\nabla_iv_j-
\nabla_jv_i$ is the vorticity tensor, $T$ is the temperature, and $R_{ij}$ 
is the Ricci tensor associated with $g_{ij}$. Note that $R_{ij}$ vanishes in 
flat space $g_{ij}=\delta_{ij}$, but keeping terms involving the curvature 
is crucial for obtaining the correct low energy expansion of $G_R$. 

 We will concentrate on the ``pure shear'' component $G_R^{xyxy}$.
For this purpose we consider a perturbation of the form $h_{xy}(z,t)$.
From the linearized Euler equation we can see that the perturbation
does not induce a shift in the density, temperature, or velocity.
This means that we can directly read off $\delta\Pi_{ij}$ from
equ.~(\ref{del_pi_fin}). We find
\be 
\label{Kubo}
 G_R^{xyxy}(\omega,k) = P -i\eta \omega +\tau_R\eta \omega^2
  - \frac{\kappa_R}{2} k^2 + O(\omega^3,\omega k^2)\, , 
\ee
which implies the familiar Kubo relation for the shear viscosity
\be 
\label{Kubo_eta}
\eta = -\lim_{\omega\to 0}\lim_{{\bf k}\to 0}
 \, \frac{d}{d\omega}\,{\rm Im}\,G^{xyxy}_R(\omega,{\bf k})
\ee  
as well as a new Kubo formula for the viscous relaxation time
\be 
\label{Kubo_tau}
\tau_R\eta = \lim_{\omega\to 0}\lim_{{\bf k}\to 0}
 \, \frac{1}{2}\frac{d^2}{d\omega^2}\,{\rm Re}\,G^{xyxy}_R(\omega,{\bf k})\, . 
\ee
This result is simpler than the corresponding formula in relativistic 
hydrodynamics \cite{Baier:2007ix}, which also involves a term proportional 
to $\kappa_R$. In the next Section we will show that in three dimensions
fluctuations lead to a $\omega^{3/2}$ term in ${\rm Re}\,G^{xyxy}_R(\omega,0)$,
see equ.~(\ref{G_R_sum}). This term is cutoff independent and completely 
fixed by $\eta$. This implies that even if fluctuations are included 
$\tau_R$ can be defined in terms of a subtracted Kubo relation. 

\section{Hydrodynamic Fluctuations}
\label{sec_fluc}

 In this section we will study the contribution of fluctuations 
to the retarded correlation function. For this purpose it is 
convenient to start from the symmetrized correlation function 
\be 
 G_S^{xyxy}(\omega,{\bf k}) = 
 \int d^3x\int dt\, e^{i(\omega t -{\bf k}\cdot{\bf x})} 
  \left\langle \frac{1}{2}
  \left\{ \Pi_{xy}(t,{\bf x}), \Pi_{xy}(0,0)\right\}
  \right\rangle \, . 
\ee
This function is related to the retarded correlator by the fluctuation
dissipation theorem. For $\omega\to 0$ we have 
\be 
\label{f-dis}
 G_S(\omega,{\bf k}) \simeq -\frac{2T}{\omega} 
  {\rm Im}\, G_R(\omega,{\bf k})\, . 
\ee
In the low frequency, low momentum limit we can use the form of the 
stress tensor in fluid dynamics, $\Pi_{xy}=\rho v_x v_y - \eta(\nabla_x 
v_y+\nabla_y v_x) + O(\nabla^2)$, and expand the hydrodynamic variables 
around their mean values, $\rho=\rho_0+\delta\rho$ etc. We will use 
the Gaussian approximation and write expectation values of products 
of fluctuating fields as products of two point functions. The ideal 
(zero derivative) terms in the stress tensor give one and two loop 
graphs involving velocity-velocity and density-density correlation 
functions. We will show in the appendix that graphs with higher derivative 
vertices, as well as graphs with additional loops are suppressed by powers 
of $\omega/(D_\eta K^2_{\it hyd})$, where $D_\eta$ is the momentum diffusion 
constant (see equ.~(\ref{Del_S_T})) and $K_{\it hyd}$ is the breakdown
scale of hydrodynamics which we will define in Sect.~\ref{sec_num}.
We will therefore concentrate on the one-loop graph
\be
\label{G_S_1l}
 G_S^{xyxy}(\omega,0) = \rho_0^2 \int \frac{d\omega'}{2\pi}
   \int \frac{d^3{\bf k}}{(2\pi)^3} 
   \Big[ \Delta_S^{xy}(\omega',{\bf k}) \Delta_S^{yx}(\omega-\omega',{\bf k})
       + \Delta_S^{xx}(\omega',{\bf k}) \Delta_S^{yy}(\omega-\omega',{\bf k})
   \Big] \, ,
\ee
where $\Delta_S^{ij}$ is the symmetrized velocity correlation function
\be 
\label{del_s_def}
 \Delta_S^{ij}(\omega,{\bf k}) = 
 \int d^3x\int dt\, e^{i(\omega t -{\bf k}\cdot{\bf x})} 
  \left\langle \frac{1}{2}
  \left\{ v^i(t,{\bf x}), v^j(0,0)\right\}
  \right\rangle \, . 
\ee
We are ultimately interested in the retarded, not the symmetrized,
correlation function. At low frequency the retarded function can be 
written as 
\bea
 G_R^{xyxy}(\omega,0) &=& \rho_0^2 \int \frac{d\omega'}{2\pi}
   \int \frac{d^3{\bf k}}{(2\pi)^3} 
   \Big[ \Delta_R^{xy}(\omega',{\bf k}) \Delta_S^{yx}(\omega-\omega',{\bf k})
       + \Delta_S^{xy}(\omega',{\bf k}) \Delta_R^{yx}(\omega-\omega',{\bf k})
\nonumber \\[0.1cm]
 & & \hspace*{0.5cm}\mbox{}
       + \Delta_R^{xx}(\omega',{\bf k}) \Delta_S^{yy}(\omega-\omega',{\bf k})
       + \Delta_S^{xx}(\omega',{\bf k}) \Delta_R^{yy}(\omega-\omega',{\bf k})
   \Big] \, , 
\label{G_R_1l}
\eea
where we have used the fluctuation dissipation relation (\ref{f-dis}).
This relation generalizes: retarded correlation functions of hydrodynamic
variables have diagrammatic expansions in terms of retarded and 
symmetrized correlation functions \cite{Ma:1976,Martin:1973zz,Hohenberg:1977ym,DeDominicis:1977fw,Khalatnikov:1983ak,Onuki:2002,Kovtun:2012rj}.

 The velocity correlation function can be decomposed into longitudinal 
and transverse parts
\be 
\label{vv_cor_lt}
 \Delta_{S,R}^{ij}(\omega,{\bf k}) = 
 \left( \delta^{ij}-\hat{\bf k}^i\hat{\bf k}^j\right) 
             \Delta_{S,R}^{T}(\omega,{\bf k})
 + \hat{\bf k}^i\hat{\bf k}^j \Delta_{S,R}^L(\omega,{\bf k})\, . 
\ee
The transverse part is purely diffusive. The symmetrized correlation 
function is \cite{Landau:smII}
\be
\label{Del_S_T}
\Delta_S^T(\omega,{\bf k}) = 
 \frac{2T}{\rho} \frac{D_\eta k^2}{\omega^2+\left(D_\eta k^2\right)^2}\, ,
\ee
where $k=|{\bf k}|$ and $D_\eta=\eta/\rho$ is the momentum diffusion
constant, also known as the kinetic viscosity. The retarded correlation 
function is given by
\be
\label{Del_R_T}
\Delta_R^T(\omega,{\bf k}) = 
 \frac{1}{\rho} \frac{-D_\eta k^2}{-i\omega+D_\eta k^2}\, .
\ee
The longitudinal correlation function can be reconstructed from 
the density-density correlation function (the dynamic structure
factor) using current conservation, $i\omega\delta\rho = i\rho{\bf k} 
\cdot {\bf v}_L$, where ${\bf v}={\bf v}_L+{\bf v}_T$ with ${\bf k}\cdot
{\bf v}_T=0$ and ${\bf k}\times {\bf v}_L=0$. We find
\bea 
\Delta_S^L(\omega,{\bf k}) &= & \frac{2T}{\rho} 
 \Bigg\{ \frac{\Gamma\omega k^2}
        {\left(\omega^2-c_s^2k^2\right)^2+\left(\Gamma\omega k^2\right)^2}
 + \left(\frac{c_p}{c_v}-1\right)\frac{1}{c_s^2}
          \frac{D_T\omega^2}
               {\omega^2+\left(D_Tk^2\right)^2} 
    \nonumber \\
 & & \hspace{0.5cm}\mbox{}
 - \left(\frac{c_p}{c_v}-1\right)\frac{1}{c_s^2}
 \frac{(\omega^2-c_s^2k^2)D_T\omega^2}
      {\left(\omega^2-c_s^2k^2\right)^2+\left(\Gamma\omega k^2\right)^2}
 \Bigg\} \, . 
\label{Del_S_L}
\eea
The first two terms have a clear physical interpretation as the contributions 
from propagating sound waves and diffusive heat transport. The third term is 
required to satisfy sum rules. This term is suppressed near the sound pole 
$\omega^2\simeq c_s^2k^2$. In equ.~(\ref{Del_S_L}) $c_s$ is the speed of 
sound, $\Gamma$ is the sound attenuation constant, $D_T=\kappa/(c_p\rho)$ is 
the thermal diffusion constant, $\kappa$ is the thermal conductivity, $c_{p}$ 
is the specific heat per unit mass at constant pressure and $c_v$ is the 
specific heat at constant volume. The sound attenuation constant is 
\be 
\label{Gamma}
 \Gamma = \frac{4}{3}\frac{\eta}{\rho}+\frac{\zeta}{\rho} 
   +\frac{\kappa}{\rho}\left(\frac{1}{c_v}-\frac{1}{c_p}\right) 
  = \frac{4}{3}\frac{\eta}{\rho}\left[
     1 + \frac{3}{4}\frac{\zeta}{\eta}
       + \frac{3}{4}\frac{\Delta c_p}{{\it Pr}} \right]\, ,
\ee
where $\Delta c_p=(c_p-c_v)/c_v$ and ${\it Pr}=(c_p\eta)/\kappa$ is the 
Prandtl number, the ratio of momentum to thermal diffusion. At high 
temperature $\Delta c_p=2/3$ and ${\it Pr}=2/3$ \cite{Braby:2010ec}, 
and at low temperature $\Delta c_p/{\it Pr}\to 0$. In the case 
of a conformal fluid the bulk viscosity vanishes, $\zeta=0$. This
implies that $\Gamma=\frac{7}{3}D_\eta$ at high temperature, and 
$\Gamma=\frac{4}{3}D_\eta$ at low temperature. 

 At low frequency and momentum the symmetrized correlation function 
can be further simplified. We illustrate the result in the case of 
the sound pole. We can write 
\be 
\label{Del_S_simp}
\Delta_S^{\it sound}(\omega,{\bf k}) \simeq \frac{\Gamma T k^2}{2\rho} 
 \left\{ \frac{1}
        {\left(\omega-c_sk\right)^2+\left(\frac{\Gamma k^2}{2}\right)^2}
     +   \frac{1}
        {\left(\omega+c_sk\right)^2+\left(\frac{\Gamma k^2}{2}\right)^2}
 \right\}\, , 
\ee
which is correct up to terms of order $\Gamma k/c_s$. The retarded
correlator is 
\be 
\label{Del_R_simp}
\Delta_R^{\it sound}(\omega,{\bf k}) \simeq \frac{\omega}{2\rho} 
 \left\{ \frac{1}
        {\omega-c_sk + i\frac{\Gamma k^2}{2}}
     +   \frac{1}
        {\omega+c_sk + i\frac{\Gamma k^2}{2}}
 \right\}\, ,
\ee
and an analogous expression holds for the sum rule term in 
equ.~(\ref{Del_S_L}).

\begin{figure}[t]
\bc\includegraphics[width=11cm]{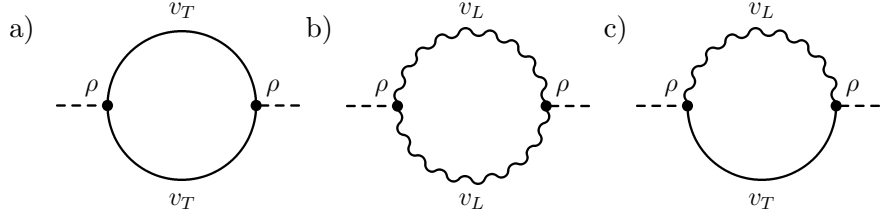}\ec
\vspace*{-4cm}\flushleft\hspace*{2cm} a)\hspace*{3.6cm}b)\hspace*{3.6cm}c)
\vspace*{3.5cm}
\caption{\label{fig_loop}
Diagrammatic representation of the leading contribution of 
thermal fluctuations to the stress tensor correlation function.
Solid lines labeled $v_T$ denote the transverse velocity correlator, 
dominated by the shear pole, and wavy lines labeled $v_L$ denote the 
longitudinal velocity correlator, governed by the sound pole and
the diffusive heat mode.}   
\end{figure}

 We can insert the decomposition of the velocity correlation function
given in equ.~(\ref{vv_cor_lt}) into the one-loop result for the retarded
correlation function, equ.~(\ref{G_R_1l}). This gives a series of terms
which correspond to the contribution from a pair of shear modes, a pair
of sound modes, a mixed shear and sound term, and finally diffusive heat 
modes, see Fig.~\ref{fig_loop}. We discuss these contributions in turn:

1. Shear modes: This is the contribution which is easiest to compute. 
The frequency integral can be done by contour integration. We find
\be 
\label{G_R_sh}
\left. G_R^{xyxy}(\omega,0)\right|_{\it shear}= -\frac{7T}{30\pi^2} 
  \int dk\, \frac{k^4}{k^2-i\omega/(2D_\eta)}\, . 
\ee
This integral is UV divergent. We regulate the divergence by introducing 
a momentum cutoff $\Lambda$. We then expand the integral in the low 
frequency regime. We get 
\be 
\label{G_R_sh_exp}
\left. G_R^{xyxy}(\omega,0)\right|_{\it shear}=- \frac{7}{90\pi^2}T\Lambda^3
  - i \omega\,\frac{7T\Lambda}{60\pi^2D_\eta} 
  + (1+i)\omega^{3/2} \frac{7T}{240\pi D_\eta^{3/2}}
  + O(\omega^{5/2})\; . 
\ee
The physical meaning of these terms can be understood by comparing 
with the Kubo relation in equ.~(\ref{Kubo}). The first term is a 
fluctuation contribution to the pressure, and the second term is 
a correction to the shear viscosity. The imaginary part of the third 
term can be viewed as a frequency dependent correction to the shear 
viscosity, and the real part is a frequency dependent contribution 
to the relaxation time which diverges as $\omega^{-1/2}$ in the 
low frequency limit. The existence of this term is sometimes 
interpreted as an indication that hydrodynamics breaks down beyond
the Navier-Stokes (one-derivative) order. 

2. Sound modes: In order to calculate the contribution from sound 
modes we use equ.~(\ref{Del_S_simp}) and (\ref{Del_R_simp}). This 
leads to two types of terms, depending on whether the real parts of
the poles of the propagators in the $\omega$-plane have the same 
or opposite sign. The contribution from terms with opposite real
parts has the same structure as the shear mode term. We get 
\be 
\label{G_R_sound_exp}
 \left. G_R^{xyxy}(\omega,0)\right|_{\it sound}= \frac{1}{90\pi^2}T\Lambda^3
  - i \omega\,\frac{T\Lambda}{30\pi^2\Gamma} 
  + (1+i)\omega^{3/2} \frac{\sqrt{2}T}{120\pi \Gamma^{3/2}}
  + O(\omega^{5/2})\; . 
\ee
The contribution from terms with real parts of the same sign is 
not infrared sensitive and does not contribute to the retarded 
correlation function at $O(\omega)$ or $O(\omega^{3/2})$.

3. Shear-sound contribution: The shear-sound contribution has the 
structure 
\be 
\label{G_R_sh_so}
 \left. G_R^{xyxy}(\omega,0)\right|_{\it sh-so} \sim 
  \int_{-\Lambda}^{\Lambda} dk \, 
  \frac{k^4}{k^2+i(\omega -c_s k)/D_s}\, , 
\ee
where $D_s=D_\eta+\Gamma/2$ and the range of the $k$-integral 
is $[-\Lambda,\Lambda]$ because of the $k\leftrightarrow -k$ symmetry 
of the sound propagator in equ.~(\ref{Del_S_simp}). We get 
\be 
\label{G_R_sh_so_exp}
 \left. G_R^{xyxy}(\omega,0)\right|_{\it sh-so} \sim 
  \frac{D_s^2\Lambda^5}{c_s^2}
   + i\omega \frac{D_s\Lambda^3}{c_s^2} + O(\omega^2)\, ,
\ee
which is suppressed relative to the pure shear and sound contributions
by a factor $D_s\Lambda/c_s\ll 1$ (see Sec.~\ref{sec_num}). We also 
note that the mixed shear-sound term does not give non-analytic terms
of the form $\omega^{3/2}$. 

4. Diffusive heat modes: The contribution of diffusive heat modes
is very similar to the shear term, but the residue of the heat 
mode is proportional to $\omega^2/c_s^2$ instead of $k^2$. In the 
diffusive regime $\omega^2\ll c_s^2 k^2$. We find
\be 
 \left. G_R^{xyxy}(\omega,0)\right|_{\it heat} \sim 
 \frac{\omega^2 D_T^2}{c_s^4} 
   \int dk\, \frac{k^2}{k^2-i\omega/(2D_T)}
 \sim \frac{\omega^2 D_T^2\Lambda^3}{c_s^4} \, , 
\ee
which is much smaller than the shear term. 

 We conclude that the main contribution arises from the pure 
shear and sound terms. We will combine these two contributions
using the approximation $\Gamma\simeq \frac{4}{3}D_\eta$, which 
corresponds to the low temperature regime. This is the more 
interesting regime because $D_\eta$ is small and the role of 
fluctuations is enhanced. We find 
\be 
\label{G_R_sum}
\left. G_R^{xyxy}(\omega,0)\right|_{\it tot}= {\it const}
  - i \omega\,\frac{17T\Lambda}{120\pi^2D_\eta} 
  + (1+i)\omega^{3/2} T
   \frac{7+\left(\frac{3}{2}\right)^{3/2}}{240\pi D_\eta^{3/2}}
  + O(\omega^2)\; . 
\ee
As noted above the $i\omega$ term is a contribution to the shear 
viscosity. This term is cutoff dependent, but the physical viscosity 
must be independent of an arbitrary cutoff. This implies that the 
bare viscosity must be cutoff dependent too, and that the cutoff
dependence of the bare viscosity is governed by a renormalization
group equation. It is important for the consistency of hydrodynamics
as an effective theory that the non-analytic $\omega^{3/2}$ term 
is not cutoff dependent, because any cutoff dependence in this
contribution cannot be absorbed into the parameters of 
hydrodynamics. 

\section{Phenomenological estimates}
\label{sec_num}

 In this section we study phenomenological implications of the 
results derived in the previous section. We have seen that the $i \omega$
term in the retarded correlation function can be combined with the 
bare shear viscosity to give a physical viscosity 
\be
\label{eta_eff} 
 \eta_{\it phys} = \eta + \frac{17}{120\pi^2}\frac{\rho T\Lambda}{\eta}\, . 
\ee
An interesting consequence of this result is the fact that the physical 
viscosity cannot be arbitrary small \cite{Kovtun:2011np}, because 
equ.~(\ref{eta_eff}) has a minimum as long as the bare viscosity is 
positive. The bare viscosity must be positive for the hydrodynamic 
expansion to be well defined. The value at the minimum depends on the 
value of the cutoff; the larger the cutoff the stronger the bound on 
$\eta$ becomes. The largest possible value of the cutoff is determined 
by the condition that the gradient expansion on which hydrodynamics
is based must be valid for all $k\lsim \Lambda$. In the following 
we will study this condition separately in the shear and sound
channel. 

1. Shear channel: Shear modes are characterized by $\omega\sim 
D_\eta k^2$. Corrections arise from higher order terms in the derivative
expansion. For non-zero frequency the leading correction is due to the 
relaxation time. We have $\omega \sim D_\eta k^2 \ll \tau_R^{-1}$. For this 
relation to be maintained for all $k<\Lambda$ we need to require that 
$\Lambda \lsim K_{\it hyd}$ with $K_{\it hyd}=(\tau_R D_\eta)^{-1/2}$. 
In kinetic theory $\tau_R=\eta/P$ \cite{Bruun:2007,Chao:2010tk,Chao:2011cy} 
and 
\be 
\label{Lam_sh}
 K_{\it hyd} \simeq \frac{1}{D_\eta}\left(\frac{P}{\rho}\right)^{1/2}\, . 
\ee

2. Sound channel: In the sound channel we have $\omega \sim c_s k \ll 
\Gamma k^2$. Using $\Gamma\simeq \frac{4}{3}D_\eta$ we find
\be 
\label{Lam_so}
 K_{\it hyd} \simeq \frac{3}{4D_\eta}
    \left(\frac{\partial P}{\partial\rho}\right)_s^{1/2}\, . 
\ee
For a weakly interacting gas $(\partial P)/(\partial\rho)_s\simeq (5P)/
(3\rho)$, and equ.~(\ref{Lam_sh}) differs from equ.~(\ref{Lam_so}) by 
a factor very close to one, $\sqrt{16/15}\simeq 1.03$. In the following 
we will use equ.~(\ref{Lam_sh}) as our estimate for the cutoff. We note 
that near a critical point the speed of sound can go to zero, and 
the contribution of sound waves is strongly suppressed relative to shear
modes. 

 It is interesting to consider the microscopic meaning of the ultraviolet 
scale $K_{\it hyd}$. In kinetic theory $\eta\sim n\bar{p}l_{\it mfp}$, 
where $\bar{p}\sim \sqrt{mT}$ is the mean momentum, and $l_{\it mfp}$ is 
the mean free path. For a weakly interacting gas $P\simeq n T$ and the 
ultraviolet scale is $K_{\it hyd}= (\rho/\eta)(P/\rho)^{1/2}\sim l^{-1}_{\it 
mfp}$. This is physically reasonable: It does not make sense to consider 
hydrodynamic fluctuations with wavelengths shorter than the mean 
free path. 

\begin{figure}[t]
\bc\includegraphics[width=7.5cm]{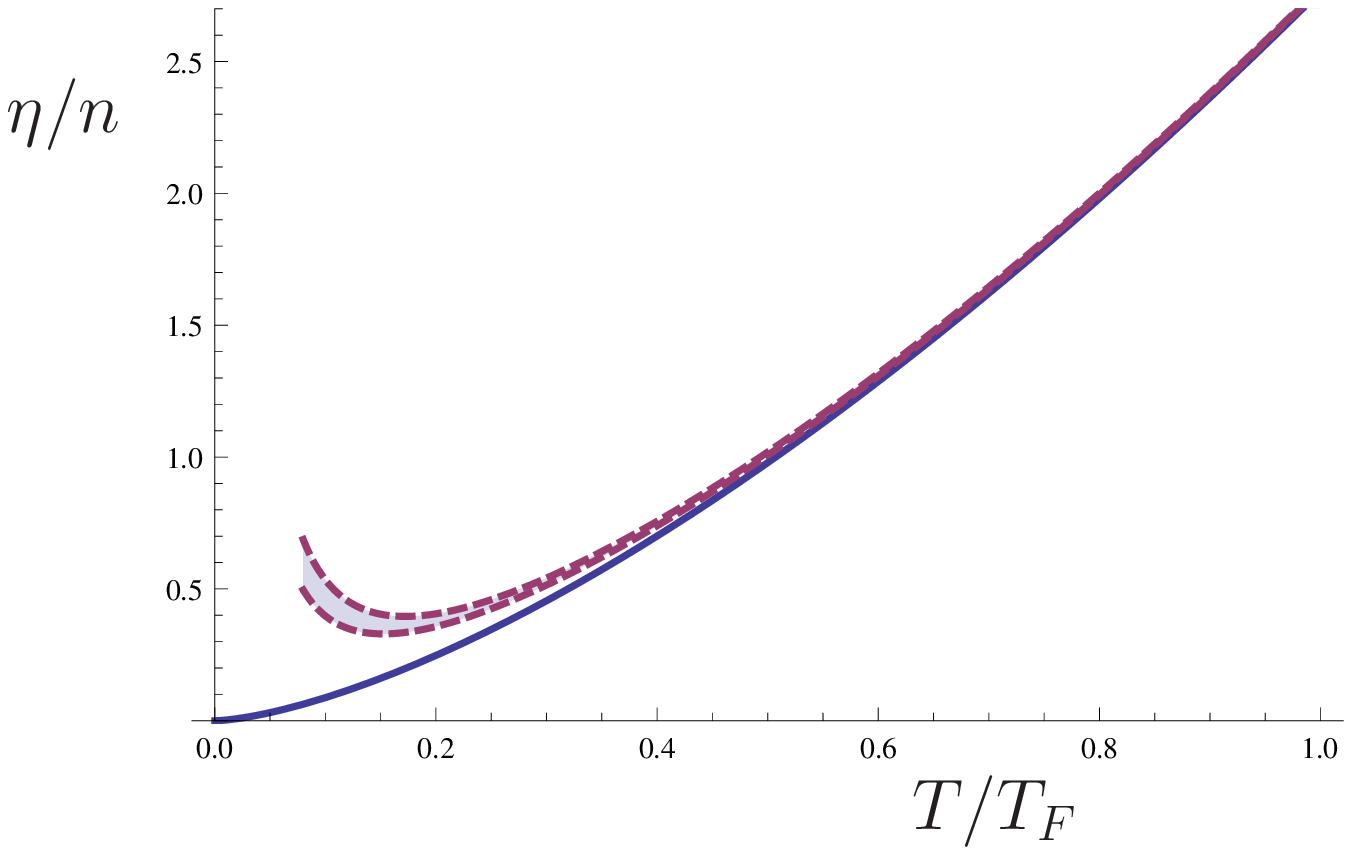}
\includegraphics[width=7.5cm]{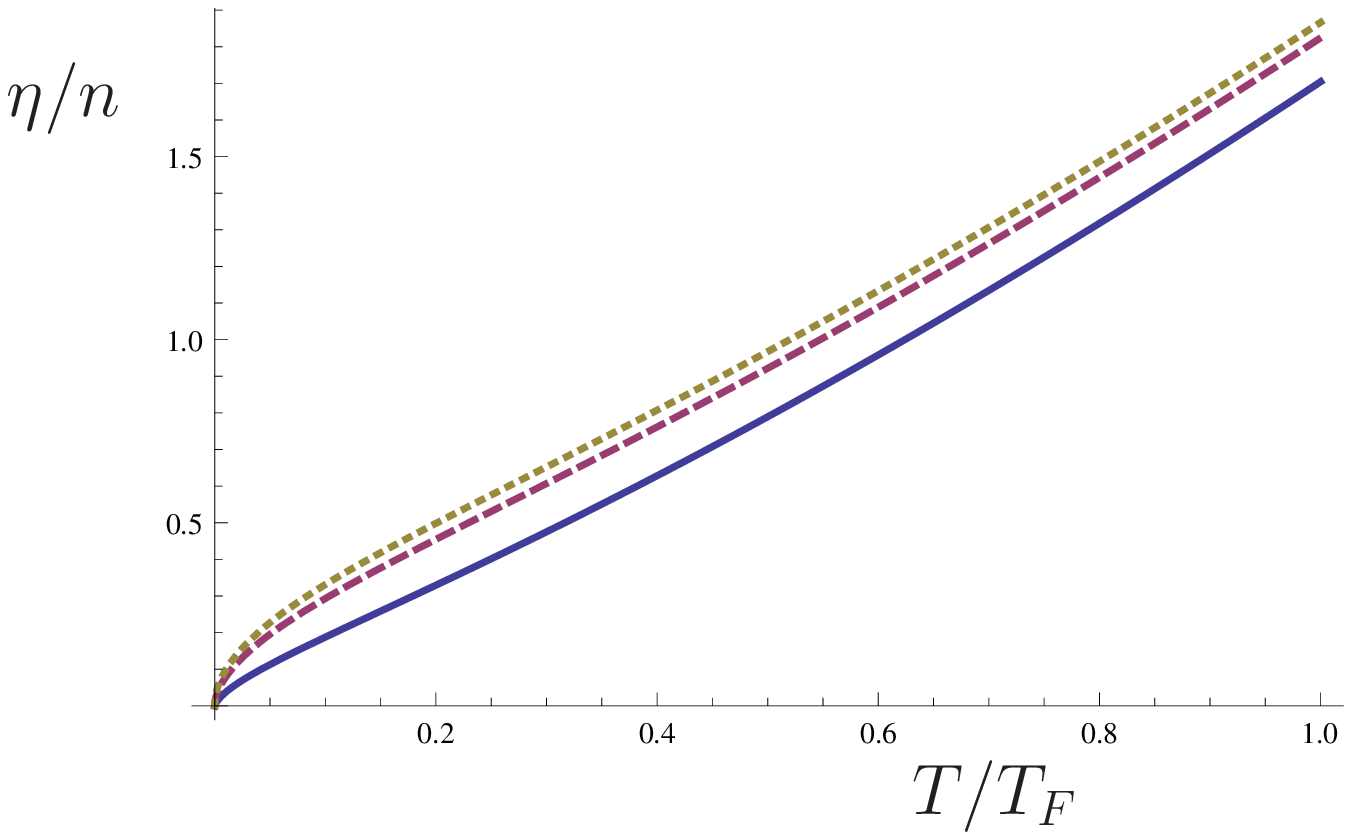}
\ec
\caption{\label{fig_eta_n}
Shear viscosity to density ratio $\eta/n$ as a function of $T/T_F$, 
where $T_F$ is the local Fermi temperature. The left panel shows $\eta/n$ 
for the unitary gas in three dimensions. The solid line is the result in 
kinetic theory and the dashed line includes fluctuations. The band shows 
the uncertainty if the cutoff is varied in the regime $\Lambda=(0.25-0.75)
K_{\it hyd}$. The right panel shows the two-dimensional gas at the crossover 
point $T_{a,2d}=T_F$. The solid line is the kinetic result. The dashed and 
dotted lines include fluctuations where we have used $\Lambda=K_{\it hyd}$ 
and $\omega$ was taken to be the frequency of the quadrupole mode 
in a harmonic trap with $N=10^4$ and $N=10^5$ particles. }   
\end{figure}

 We can illustrate this result further by using the leading order 
kinetic theory result as an estimate for the bare viscosity. This is 
consistent because kinetic theory takes into account effects at distances 
$l\lsim l_{\it mfp}$ but, unless stochastic forces are included, it does not 
take into account fluctuations on length scales $l\gsim l_{\it mfp}$. The 
kinetic theory result is \cite{Massignan:2004}
\be 
\label{eta_kin}
 \eta = \frac{15}{32\sqrt{\pi}} \, (mT)^{3/2} \, .
\ee
In Fig.~\ref{fig_eta_n} we show the bare viscosity and the physical viscosity 
including the effects of fluctuations. The band shows the uncertainty if 
the cutoff is varied in the regime $\Lambda=(0.25-0.75) K_{\it hyd}$. We observe 
that the viscosity has a minimum $\eta/n\simeq 0.5$ at a temperature $T\simeq 
0.2T_F$, close to the critical temperature $T_c=0.167(13)T_F$ \cite{Ku:2012}. 
Note that the increase of the shear viscosity at low temperature does not 
imply a breakdown of the hydrodynamic expansion: In this regime the 
one-loop graph is large compared to the bare viscosity, but the power
counting discussed in App.~\ref{sec_app} ensures that graphs with more 
loops are suppressed.

 The increase of the shear viscosity in the low temperature regime is related 
to a non-analytic frequency dependence of $\eta(\omega)=-{\rm Im}\,G^{xyxy}_R
(\omega,{\bf k}\!=\!0)$.  Equ.~(\ref{G_R_sum}) implies that for small 
$\omega$
\be 
\label{eta_non_analyt}
\eta(\omega) = \eta - \sqrt{\omega}\,T\, 
  \frac{7+\left(\frac{3}{2}\right)^{3/2}}{240\pi D_\eta^{3/2}}\, . 
\ee
The width of the non-analytic structure in the spectral function can 
be estimated by assuming that the fluctuation term in the physical 
shear viscosity, the second term in equ.~(\ref{eta_eff}), is due 
to the non-analytic term in the spectral function. This assumption 
leads to $\Delta\omega \simeq 0.3 T(n/\eta)$, where we have used 
$\Lambda\simeq 0.5 K_{\it hyd}$. 

\section{The bound on $\eta/s$}
\label{sec_bound}

\begin{figure}[t]
\bc\includegraphics[width=8.5cm]{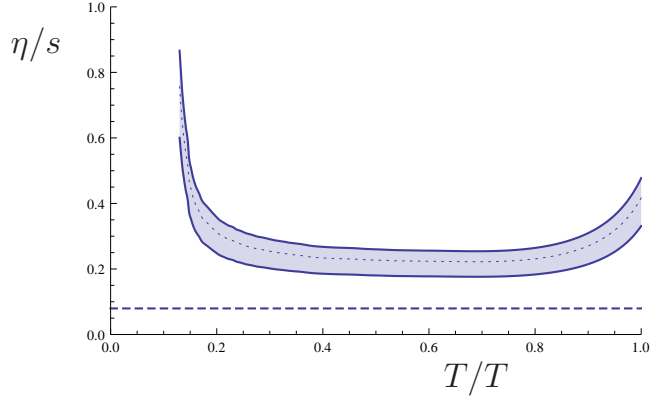}\ec
\caption{\label{fig_eta_s}
Bound on the shear viscosity to entropy density ratio $\eta/s$ as a 
function of $T/T_F$, where $T_F$ is the local Fermi temperature. This
figure shows the bound given in equ.~(\ref{eta_bound}) evaluated 
using measurements of thermodynamic properties reported in \cite{Ku:2012}.
The band around the dotted line shows the sensitivity to variations in 
the cutoff in the range $\Lambda=(0.5\pm 0.25)K_{\it hyd}$. The dashed 
line shows the string theory bound $\eta/s=1/(4\pi)$.}   
\end{figure}

 The model discussed in the previous section shows that even if the 
bare viscosity goes to zero the physical viscosity always finite. 
In this section we show that there is a lower bound on $\eta/s$
which does not depend on assumptions about the temperature dependence
of the bare viscosity. Equ.~(\ref{eta_eff}) implies 
\be 
\label{eta_phys}
  \left(\frac{\eta}{s}\right)_{\it phys} 
  = \frac{\eta}{s} 
   + \frac{17}{\sqrt{2}\, 80}
      \bigg(\frac{s}{\eta}\bigg)^2
      \bigg(\frac{T}{T_F}\bigg)^{3/2}
      \bigg(\frac{n}{s}\bigg)^3
      \bigg(\frac{P}{nT}\bigg)^{1/2}
      \bigg(\frac{\Lambda}{K_{\it hyd}}\bigg)\, . 
\ee
Minimizing this expression with respect to the bare viscosity we find
\be 
\label{eta_bound}
 \left(\frac{\eta}{s}\right)_{\it phys}
 \gsim 
   1.005  \bigg(\frac{T}{T_F}\bigg)^{1/2}
   \bigg(\frac{n}{s}\bigg)
   \bigg(\frac{P}{nT}\bigg)^{1/6}
   \bigg(\frac{\Lambda}{K_{\it hyd}}\bigg)^{1/3}\, . 
\ee
This expression depends on the thermodynamic quantities $s/n$ and
$P/(nT)$, but we note that the bound on $(\eta/s)_{\it phys}$
always has a minimum at some temperature of order $T_c$. To see this
we note that $(s/n)\sim T^3$ and $P/(nT)\sim n^{1/3}/(mT)$ for 
$T\ll T_c$, whereas $(s/n)\sim \log(T)$ and $P/(nT)\sim 1$ for 
$T\gg T_c$. This implies that the bound scales as $T^{-16/6}$ at 
low $T$, and as $T^{1/2}/\log(T)$ at high $T$. In order to be more 
quantitative we have evaluated equ.~(\ref{eta_bound}) using the 
equation of state measured by the MIT group \cite{Ku:2012}, see 
Fig.~\ref{fig_eta_s}. 

 The remaining uncertainty is related to the value of the cutoff.
The validity of hydrodynamics implies that $\Lambda$ cannot be 
much smaller than $K_{\it hyd}$. This statement can be quantified 
by analyzing the data on collective modes published by the Duke
group \cite{Kinast:2005}. For the specific trap parameters 
used in that experiment the radial breathing mode was found 
to behave hydrodynamically for temperatures $T\lsim 0.8T_F$, see
\cite{Schaefer:2009px}. This implies that\footnote{We have used 
the trap parameters given in Sect.~\ref{sec_trap} below.} 
$\omega/(D_\eta\Lambda^2)\gsim \omega/(D_\eta K_{\it hyd}^2)\simeq 
0.5$. In order for the expansion parameter to satisfy $\omega/(D_\eta
\Lambda^2)<1$ the cutoff $\Lambda$ cannot be much smaller than 
$K_{\it hyd}$. On the other hand, $\Lambda$ also cannot be much 
bigger than $K_{\it hyd}$ because then higher loop corrections are 
not suppressed, see App.~\ref{sec_app}. In Fig.~\ref{fig_eta_s} 
we have used $\Lambda/K_{\it hyd}=0.5\pm 0.25$. We note that the
bound on $\eta/s$ scales as $(\Lambda/K_{\it hyd})^{1/3}$, and is
only weakly sensitive to the uncertainty in the cutoff. 

 We obtain a fairly broad minimum $(\eta/s)_{\it phys}\gsim 0.2$ 
in the regime $T/T_F\sim (0.3-0.9)$. The bound on $\eta/s$ becomes 
large as $T\to 0$ and $T\to\infty$, consistent with the expectation
from kinetic theory which predicts $\eta/s\sim (T_F/T)^8$ at low 
temperature and $\eta/s\sim (T/T_F)^{3/2}/\log(T/T_F)$ at high 
temperature \cite{Massignan:2004,Rupak:2007vp,Manuel:2011ed}. The 
bound is compatible with the experimental results reported by Cao et 
al.~\cite{Cao:2010wa} and the T-matrix calculation of Enss et 
al.~\cite{Enss:2010qh}, but lager than the Path Integral Monte Carlo 
(PIMC) results obtained by Wlazlowski et al.~\cite{Wlazlowski:2012jb}. 
These authors find $(\eta/s)_{\it min}\sim 0.2$ at temperatures $0.15T_F
\lsim T\lsim 0.25 T_F$. A possible reason for the discrepancy is that 
for the lattice spacing used in \cite{Wlazlowski:2012jb} one cannot 
resolve the non-analytic behavior of the spectral function given 
in equ.~(\ref{eta_non_analyt}).

\section{Two dimensional systems}
\label{sec_2d}

 It is interesting to consider the role of fluctuations in two 
dimensional systems. The results in Sect.~\ref{sec_fluc} are 
easily generalized to two spatial dimensions. Aside from the 
obvious substitution $d^3k/(2\pi)^3\to d^2k/(2\pi)^2$ the only 
difference is that in two spatial dimensions the shear contribution
to the sound attenuation constant is $\Gamma=D_\eta$ instead of 
$\Gamma=\frac{4}{3}D_\eta$. In both two and three dimensions the 
dominant contribution to $G_R^{xyxy}$ arises from the one loop 
diagram involving either a pair of shear modes or a pair of 
sound modes. In $d=2$ the loop integral is logarithmically 
divergent and 
\be 
\label{G_R_2d}
\left. G_R^{xyxy}(\omega,0)\right|_{\it tot}= {\it const}
  - i \omega\,\frac{T}{16\pi D_\eta} \left[
     \log\left(\frac{\sqrt{2}D_\eta\Lambda^2}{\omega}\right)
      + i\frac{\pi}{2}\right]\, . 
\ee
The imaginary part can be interpreted as a correction to the 
shear viscosity. We find
\be 
\label{eta_phys_2d}
 \eta_{\it phys}= \eta + \frac{mT}{16\pi}\frac{n}{\eta}
  \log\left(\frac{\sqrt{2}\Lambda^2}{m\omega}\frac{\eta}{n}\right)
\ee
which diverges logarithmically as $\omega\to 0$. This divergence is 
well known \cite{Ernst:1971,Forster:1977,Khalatnikov:1984}, and it 
has been observed in molecular dynamics and lattice gas simulations 
\cite{Alder:1967,Kadanoff:1989}. To the best of our knowledge it has 
not been observed experimentally. Equ.~(\ref{eta_phys_2d}) shows that 
the effect is large in systems that have a small value of the bare 
shear viscosity. 

 The shear viscosity of a dilute two-dimensional Fermi gas was
recently computed in \cite{Schafer:2011my,Bruun:2011}. The result is 
\be 
\label{eta_2d}
\eta = \frac{mT}{2\pi^2}
 \left(\left[\log\left(\frac{5T}{2T_{a,2d}}\right)\right]^2
+\pi^2\right)\, ,
\ee
where $T_{a,2d}=1/(ma_{2d})^2$ and $a_{2d}$ is the scattering length in 
two dimensions \cite{Randeria:1990}. In two dimensions there is no scale 
invariant fluid except in the non-interacting limit $a_{2d}\to 0$. The 
most strongly correlated fluid corresponds to $T_{a,2d}\simeq T_F$, which 
implies that the dimer binding energy is equal to the Fermi energy. The 
viscosity of the two dimensional gas was recently studied by measuring the 
damping of the quadrupole mode \cite{Vogt:2011} in a harmonic potential. 
The frequency of  this mode is $\omega=\sqrt{2}\omega_\perp$, where 
$\omega_\perp$ is the two dimensional oscillator frequency. The 
confinement frequency sets the scale for the Fermi temperature of the 
trap, $T_F^{\it trap}=N^{1/2}\omega^\perp$. We can use these relations 
to translate the frequency dependence of the shear viscosity into the 
dependence on the number of particles. In Fig.~\ref{fig_eta_n} we show 
the kinetic theory result as well as the viscosity with fluctuations 
included for two different values of the particle number. The main 
dependence on $N$ is of the form $(\eta/n)_{\it phys} \sim \frac{1}
{16\pi}\log(N)$. We observe that fluctuations make a significant 
contribution to the shear viscosity, but the logarithmic divergence 
with $N$ is fairly slow, and one will need significantly larger numbers 
of particles than what is available in current experiments ($N\simeq 
4\cdot 10^3$ in \cite{Vogt:2011}) to see the effect clearly. 

\section{Trapped atomic gases}
\label{sec_trap}

 In this section we will try to make contact with experiments
that study the damping of collective modes in trapped Fermi 
gases. We are interested in the question whether it is possible 
to establish the role of hydrodynamic fluctuations by studying
the scaling of the damping constant with temperature or particle 
number. A review of the hydrodynamic theory of collective modes
can be found in \cite{Schaefer:2009px}.

\begin{figure}[t]
\bc\includegraphics[width=11cm]{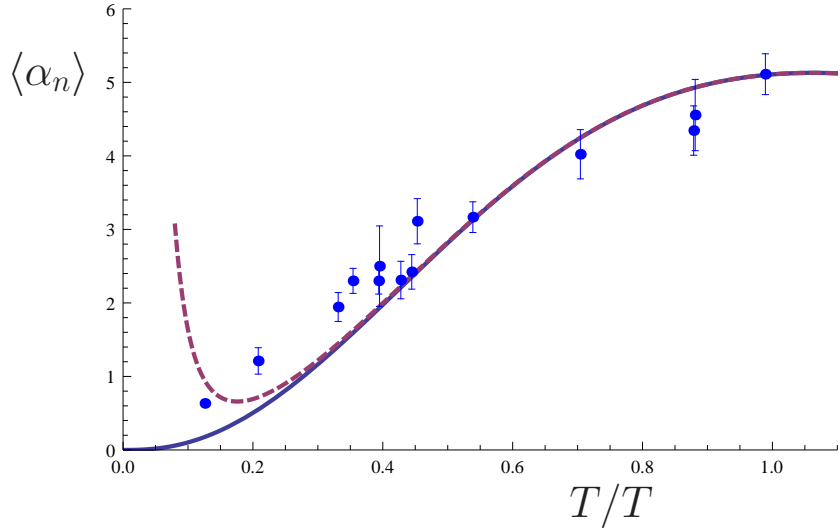}
\ec
\caption{\label{fig_eta_trap}
Trap averaged shear viscosity to density ratio $\langle\alpha_n\rangle$. 
We show $\langle\alpha_n\rangle$ as a function of $T/T_F^{\it trap}$, where
$T_F^{\it trap}=(3\lambda N)^{1/3}\omega_\perp$ is the Fermi temperature of 
the trap. We have chosen $N=2\cdot 10^5$ and $\lambda=0.045$ as in 
\cite{Kinast:2005}. The solid line shows the kinetic theory result, 
the dashed line includes fluctuation corrections to the shear viscosity.
The data are from \cite{Cao:2011fh}, which is a reanalysis of the
results reported in \cite{Kinast:2005}. We do not show data in 
the superfluid regime $T\ll T_c$. }   
\end{figure}

 Consider a trapped gas with $N$ particles in a harmonic potential 
with trapping frequencies $\omega_x=\omega_y=\omega_\perp$ and $\omega_z
=\lambda\omega_\perp$. In a typical experiment $N=(10^5-10^6)$ and $\lambda
=(0.02-0.05)$ \cite{Kinast:2004b,Bartenstein:2004,Kinast:2005,Altmeyer:2006}. 
The transverse breathing mode has a frequency $\omega=\sqrt{10/3}\,
\omega_\perp$ and the damping constant is 
\be 
 \Gamma_{\it br} =  \frac{\langle \alpha_n\rangle}{(3N\lambda)^{1/3}}
    \frac{\omega_\perp}{(E_0/[N\epsilon_F])}\, , 
\ee
where $E_0$ is the total (potential and internal) energy of the trapped
gas, $\epsilon_F = (3N\lambda)^{1/3}\omega_\perp$ is the Fermi energy
of the trapped system, and $\langle\alpha_n\rangle=\frac{1}{N}\int 
d^3x\,\eta(x)$ is the trap average of the shear viscosity. Taking into 
account relaxation time effects we have 
\be 
\langle \alpha_n\rangle = \frac{1}{N} 
  \int d^3x\, \frac{\eta(x)}{1+\omega^2\tau_R(x)^2}\, . 
\ee
We take the bare shear viscosity from kinetic theory, equ.~(\ref{eta_kin}),
and compute the physical viscosity from equ.~(\ref{eta_phys}). We note 
that the bare viscosity only depends on $T$, which is independent of 
the position in the trap. The fluctuation term is largest at the center 
of the trap. We also use kinetic theory to determine the bare 
relaxation time, $\tau_R = \eta/P$, and use equ.~(\ref{G_R_sum}) 
to determine the physical relaxation time. This corresponds to
\be
(\tau_R\eta)_{\it phys} = \tau_R\eta + 
  \frac{\left[7+\left(\frac{3}{2}\right)^{3/2}\right]T}
       {240\pi\omega^{1/2}}
  \left(\frac{\rho}{\eta}\right)^{3/2}\, . 
\ee
We note that the bare relaxation time is inversely proportional 
to the local pressure and depends on the position in the trap. 
In particular, $\tau_R\eta$ is large in the dilute part of the 
cloud. Fluctuations, on the other hand, increase the relaxation
time near the center of the trap. 

In the following we will use the high temperature approximation
for the density of the cloud. This is consistent with using kinetic
theory for the bare shear viscosity and relaxation time. It also
provides a very accurate description of the tail of the density 
distribution at essentially all temperatures. We have 
\be 
\label{n_3d}
 n(x) = N \left(\frac{m\bar{\omega}^2}{2\pi T}\right)^{3/2}
   \exp\left(-\sum_i \frac{m\omega_i^2x_i^2}{2T}\right)\, , 
\ee
where $\bar\omega=(\omega_\perp^2\omega_z)^{1/3}$. Results for $\langle
\alpha_n\rangle$ as a function of $T/T_F^{\it trap}$ with $T_F^{\it trap}
=(3N\lambda)^{1/3}\omega_\perp$ are shown in Fig.~\ref{fig_eta_trap}. 
We have used $N=2\cdot 10^5$ and $\lambda=0.045$ as in the experiment 
of Kinast et al.~\cite{Kinast:2005}. The solid line shows the result 
using kinetic theory for $\eta$ and $\tau_R$, and the dashed line 
includes the fluctuation term in $\eta$. We find that for the parameters 
considered here corrections to the relaxation time are very small. 

We observe that kinetic theory describes the data for $T\gsim 0.4 
T^{\it trap}_F$ well. Fluctuations are important for $T\lsim 0.2 
T^{\it trap}_F$, leading to a minimum in $\langle\alpha_n\rangle$. 
We note that the critical temperature is $T_c\simeq 0.2 T^{\it trap}_F$
\cite{Luo:2009}, and we do not expect the theory used in this section,
which is based on kinetic theory in the dilute limit, to reproduce
experiment for $T\ll T_c$.  It was recently suggested that the data
in this regime are dominated by the transition from hydrodynamic
to ballistic behavior \cite{Mannarelli:2012su}. 

 Finally, we have looked at the role of fluctuations in the experiment 
of Vogt et al.~\cite{Vogt:2011}. In this experiment the damping of two 
dimensional quadrupole mode was measured for $N=4\cdot 10^3$. The 
dependence on the scattering length was studied for $\log(k_Fa_{2d})>0$ 
at $T/T_F=0.48$, and the temperature dependence was studied in the range 
$T/T_F^{\it trap}=(0.3-0.8)$ for $\log(k_Fa_{2d})=(2.7-42)$. We find
that for this range of parameters the role of fluctuations is always
small. Fluctuations lead to a significant enhancement of the damping
constant if $\log(k_Fa_{2d})\sim 0$ and $N\gsim 10^5$. This 
enhancement grow as $\log(N)$, but the logarithmic growth is 
difficult to disentangle from a $\log(N)$ term related to the 
dilute corona, see \cite{Schafer:2011my}.

\section{Conclusions and outlook}
\label{sec_out}

 We have studied the role of hydrodynamics fluctuations in the dilute 
Fermi gas. Our main findings are:

\begin{enumerate}
\item Hydrodynamic fluctuations imply the existence of a minimum in 
the shear viscosity. The physical origin of the minimum is the contribution 
of shear and sound modes to momentum transport. If the bare viscosity is 
small, then sound and shear modes are weakly damped and the contribution 
of hydrodynamic modes to momentum transport is large. The magnitude of 
the minimum shear viscosity is weakly dependent on the cutoff scale 
of the hydrodynamic description. Allowing for a factor of two uncertainty 
in our estimate of $\Lambda$ we find $\eta/s\gsim 0.2$. The uncertainty 
can be reduced by computing higher loop corrections. Our estimate is 
consistent with trap averaged measurements of $\eta/s$ reported in
\cite{Cao:2010wa}, but not with recent lattice calculations
\cite{Wlazlowski:2012jb}. 

\item Contrary to the proposed string theory limit $\eta/s\geq 1/(4\pi)$ 
the bound is not universal. It depends on thermodynamic properties and 
the breakdown scale of hydrodynamics. We note that the bound itself
is purely classical, $\hbar$ only enters though thermodynamic quantities. 
Ignoring numerical factors we have $\eta/s\gsim (n/s)(mT/n^{2/3})^{1/2}
(P/(nT))^{1/6}$. At large temperature the ratio $n/s$ depends weakly 
on $T$ and the bound grows as $T^{1/2}$. At low $T$ the entropy per 
particle increases sharply when the system reaches quantum degeneracy, 
which corresponds to $mT\sim \hbar^2 n^{2/3}$. This implies that at the 
minimum $\eta/s\sim \hbar$. 

\item Fluctuations cause a $1/\sqrt{\omega}$ divergence of the 
viscous relaxation time in three dimensions, and a $\log(\omega)$
divergence of the shear viscosity in two dimensions. These effects 
are independent of the cutoff and only depend on the value of the 
bare shear viscosity. The existence of non-analytic terms implies 
that, strictly speaking, the two dimensional Navier-Stokes equation 
as well as the three dimensional second order (Burnett) equations
are not consistent unless fluctuating forces are taken into account. 
We note, however, that real flows that can be studied in experiment
involve finite frequencies or time scales, and fluctuating forces
may not be important.

\item We have studied the importance of fluctuations for the damping 
of trapped Fermi gases. The corrections are generally small for
the conditions that have been experimentally investigated. A 
possible exception is the three dimensional unitary gas near $T_c$. 
In this case fluctuations may lead to enhanced damping\footnote{Note 
that we have not considered the role of critical fluctuations. The 
superfluid transition is described by model F in the classification 
of Hohenberg and Halperin \cite{Hohenberg:1977ym}. This model does 
not contain direct couplings between the order parameter and the 
momentum density, and the calculation discussed in our work is not 
directly affected by critical fluctuations.}. Fluctuations lead
to a $\log(N)$ divergence in the damping constant of the two 
dimensional Fermi gas, but this effect is difficult to observe 
unless the number of particles is varied by more than an order 
of magnitude. 

\end{enumerate}

 There are a number of interesting formal questions that we have 
not studied in this paper. In order to study higher order corrections 
it is useful to start from an effective action for hydrodynamic 
fluctuations. This could be done using the  methods developed in 
\cite{Khalatnikov:1983ak,Kovtun:2012rj}. The effective action might 
also be useful for studying the renormalization group evolution in 
more detail. In this work we have simply assumed that the bare shear 
viscosity can be computed in kinetic theory. It would be desirable 
to provide a more rigorous justification for this approximation by 
studying the matching between kinetic theory and hydrodynamics. 
Finally, in this work we have restricted ourselves to studying the 
effect of fluctuations on the damping of collective modes. In 
this case it is straightforward to take into account a frequency
dependent shear viscosity and relaxation time. If one considers
hydrodynamic flows that are not periodic in time, for example
the elliptic flow experiment described in \cite{Cao:2010wa},
one has to solve the hydrodynamic equations with fluctuating 
forces. This method has been studied in the context of 
microfluidic systems, see \cite{Balboa:2011} and references therein.

 We have shown that hydrodynamic fluctuations are important if 
the bare viscosity is small in the low temperature limit. The main 
physical question is whether this scenario is realized in the 
unitary Fermi gas near $T_c$, or whether other effects, like pairing
correlations, a pseudo-gap or phonons are more important 
\cite{Enss:2010qh,Bruun:2007b,Guo:2010dc,Guo:2010xu}. This question 
is probably difficult to address experimentally, but it can be studied 
by analyzing the spectral function of the stress tensor. For this 
purpose it is necessary to construct models of the spectral function 
that include fluctuations. These models can be confronted with quantum 
Monte Carlo data, for example the recent work of Wlazlowski et 
al.~\cite{Wlazlowski:2012jb}.

 Acknowledgments: We thank Paul Romatschke for useful discussions, 
in particular for bringing equ.~(\ref{eta_non_analyt}) to our 
attention. After the first version of this paper was submitted
related work appeared in \cite{Romatschke:2012}. This work was 
supported in parts by the US Department of Energy grant DE-FG02-03ER41260. 

\appendix
\section{Stress tensor correlation function in hydrodynamics}
\subsection{Low energy expansion}
\label{sec_app}

 In this section we provide some additional details regarding the 
low energy expansion of the retarded stress tensor correlation function. 
Our starting point is the symmetrized correlation function
\be 
\label{G_S_Pi_xy}
 G_S^{xyxy}(\omega,{\bf k}) = 
 \int d^3x\int dt\, e^{i(\omega t -{\bf k}\cdot{\bf x})} 
  \left\langle \frac{1}{2}
  \left\{ \Pi_{xy}(t,{\bf x}), \Pi_{xy}(0,0)\right\}
  \right\rangle \, . 
\ee
We use the expression for the stress tensor in hydrodynamics 
and expand in small fluctuations, $\delta\rho,\delta T,v_i$, and 
in the number of derivatives. We get
\be 
\label{del_pi_xy}
 \Pi_{xy} = \rho_0 v_xv_y 
  -\eta_0\left(\nabla_x v_y+\nabla_y v_x\right)
  - \left[\left(\frac{\partial\eta}{\partial\rho}\right)_{\!T}\delta\rho
     +\left(\frac{\partial\eta}{\partial T}\right)_{\!\rho}\delta T \right]
    \left(\nabla_x v_y+\nabla_y v_x\right) + \ldots \, 
\ee
where we have dropped terms of order $O(\delta^3)$ and $O(\nabla^2)$
(note that $v_i$ is a quantity of $O(\delta)$). The 
diagrammatic expansion can be derived by inserting equ.~(\ref{del_pi_xy})
into equ.~(\ref{G_S_Pi_xy}) and factorizing the expectation value
into pairs of fluctuating fields. The retarded correlation function
is obtained by replacing one of the symmetrized functions by a 
retarded function.
In Section \ref{sec_fluc} we computed the one loop diagram that arises
from the first term in equ.~(\ref{del_pi_xy}). This term has no spatial 
derivatives, and we find a contribution of the form
\be 
\label{G_R_w_exp}
 G_R^{xyxy}(\omega,0)\sim T\Lambda^3 
 \left\{ 1 + c_1 \frac{\omega}{D_\eta\Lambda^2} 
           + c_{3/2} \frac{\omega^{3/2}}{D_\eta^{3/2}\Lambda^3}
           + \ldots \right\},
\ee
where $c_1$ and $c_{3/2}$ are numerical constants. We observe that 
the low energy expansion involves powers of $\omega/(D_\eta\Lambda^2)$.

\begin{figure}[t]
\bc\includegraphics[width=11cm]{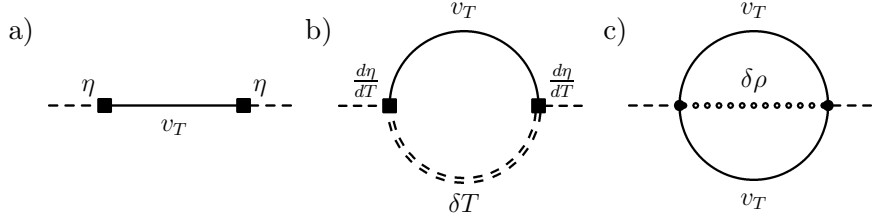}\ec
\vspace*{-4cm}\flushleft\hspace*{2.1cm} a)\hspace*{3.6cm}b)\hspace*{3.6cm}c)
\vspace*{3.5cm}
\caption{\label{fig_loop_2}
Diagrammatic representation of higher order fluctuation contributions 
to the stress tensor correlation function. Solid lines labeled $v_T$ 
denote the transverse velocity correlator, wavy lines labeled $v_L$ 
denote the longitudinal velocity correlator. Temperature fluctuations
are shown as double dashed lines, and density fluctuations are 
shown as dotted lines. Vertices shown as dots contain no derivatives,
whereas vertices labeled by squares contain spatial derivatives.}   
\end{figure}

The second term contains spatial derivatives, and it gives rise to a tree 
diagram (Fig.~\ref{fig_loop_2}a) which vanishes as $k\to 0$. The third 
term in equ.~(\ref{del_pi_xy}) involves derivatives of the shear viscosity 
with respect to $\rho$ and $T$. In kinetic theory $[(\partial\eta)/(\partial 
\rho)]_T$ vanishes at leading order in $n\lambda_{\it dB}^3$, where $\lambda_{\it 
dB}$ is the de Broglie wave length. The dominant term therefore involves 
fluctuations of the temperature, $G_S\sim [(\partial\eta)/(\partial T)]^2 
\langle \delta T \delta T\rangle \langle \nabla_xv_y \nabla_x v_y\rangle$ 
(plus permutations $x\to y$), see Fig.~\ref{fig_loop_2}b. This is a one-loop 
graph with vertices that contain one spatial derivative. For non-zero external 
momenta this graph gives a contribution to $G_R$ which is suppressed by 
$k^2/K_{\it hyd}^2$ relative to equ.~(\ref{G_R_w_exp}). For zero external 
momentum the diagram has power divergences which contribute at $O(\Lambda^2
/K_{\it hyd}^2)$. This is not small if $\Lambda\sim K_{\it hyd}$, but power 
divergences can be absorbed into the transport coefficients\footnote{It 
is well known that power counting is not manifest in effective field 
theories regularized by a momentum cutoff \cite{Burgess:1998ku}. This 
problem can be circumvented using dimensional regularization (DR), 
which automatically eliminates all power divergent terms. In our 
context this is a disadvantage because we find that the leading one 
loop divergence in the stress tensor correlation function represents
an important physical effect. The linear divergence is preserved
in a modified version of dimensional regularization called ``power
divergence subtraction'' (PDS) \cite{Kaplan:1998tg}, which keeps the
pole corresponding to the logarithmic divergence in two spatial 
dimensions.}.

 The leading no-derivative contribution of order $\delta^3$ in the 
stress tensor is of the form $\Pi_{xy}=(\delta\rho) v_x v_y$. This 
term generates the two-loop diagram shown in Fig.~\ref{fig_loop_2}c.
The extra loop integral involves three powers of momentum, and the 
extra propagator is proportional to $1/\rho$. As a consequence, 
the graph is suppressed by $\Lambda^3/k_F^3$. This can be written
as $(\Lambda/k_F)^3\lsim (K_{\it hyd}/k_F)^3\simeq 1/(nl_{\it mfp}^3)$,
where we have used the kinetic theory estimate $\eta\simeq n
(mT)^{1/2}l_{\it mfp}$. 

 In summary, the hydrodynamic expansion of the correlation function
$G_R^{xy}(\omega,{\bf k})$ involves powers of $k/K_{\it hyd}$, where 
$K_{\it hyd}=(P/\rho)^{1/2}D_\eta^{-1}\sim l_{\it mfp}^{-1}$ is the 
breakdown scale defined in Section \ref{sec_num}. Since $G_R^{xy}
(\omega,{\bf k})$ is diffusive the frequency scales as $\omega\sim 
D_\eta k^2$ and the frequency expansion involves $(\omega/[D_\eta
K_{\it hyd}^2])$. Powers of $k/K_{\it hyd}$ arise from higher 
derivative terms in the currents or from loop graphs. Loop
graphs are additionally suppressed by $(K_{\it hyd}/k_F)^3\simeq 
1/(nl_{\it mfp}^3)$. Loop graphs are important because they lead
to non-analytic effects, logarithms and fractional powers of 
$\omega$ and $k^2$. For low viscosity fluids the mean free
path is short, $nl_{\it mfp}^3\sim 1$, and there is no suppression
of loops relative to gradient terms. 

\subsection{Contact terms}
\label{sec_contact}

 The symmetrized correlation function contains a contact term
which can be determined using the fluctuation-dissipation 
theorem \cite{Landau:smII,Fox:1970}
\be
\label{G_S_cont}
\left\langle \frac{1}{2}
  \left\{ \Pi_{xy}(t,{\bf x}), \Pi_{xy}(t',{\bf x}')\right\}
 \right\rangle = 2\eta T 
   \delta(t-t')\delta({\bf x}-{\bf x}')\, . 
\ee
In frequency space this gives the contribution of the bare shear 
viscosity to the retarded correlation function, $G_R=i\omega\eta$. The 
result therefore justifies combining the bare and loop corrections as 
in equ.~(\ref{eta_eff}). The contact term can also be obtained using 
the velocity correlation function combined with the conservation 
laws \cite{Kovtun:2011np}. Momentum conservation implies $\nabla_i
\Pi^{ij}= -\frac{\partial}{\partial t} (\rho v^j)$ and 
\be 
 k_x^2 G^{xyxy}_S(\omega, k_x) = \omega^2 \Delta^{yy}_S(\omega,k_x)\, . 
\ee
Using the explicit form of the velocity correlation function 
we find
\be 
  G^{xyxy}_S(\omega,k_x) = 2\eta T \left\{ 1 - 
     \frac{(D_\eta k_x^2)^2}{\omega^2+(D_\eta k_x^2)^2} \right\}\, , 
\ee
which contains the contact term $2\eta T$.



\begin{thebibliography}{20}

\bibitem{Garcia:2008}
L.~S.~Garcia-Colina, R.~M.~Velascoa, F.~J.~Uribea, 
``Beyond the Navier-Stokes equations: Burnett hydrodynamics''
Phys.\ Rep.\ {\bf 465} 149 (2008).

\bibitem{Chao:2011cy}
J.~Chao, T.~Sch\"afer,
``Conformal symmetry and non-relativistic second order fluid dynamics,''  
Annals Phys.\  {\bf 327}, 1852 (2012)
[arXiv:1108.4979 [hep-th]].

\bibitem{Romatschke:2009im}
P.~Romatschke,
``New Developments in Relativistic Viscous Hydrodynamics,''
arXiv:0902.3663 [hep-ph].

\bibitem{Hohenberg:1977ym} 
P.~C.~Hohenberg and B.~I.~Halperin,
``Theory of Dynamic Critical Phenomena,''
Rev.\ Mod.\ Phys.\  {\bf 49}, 435 (1977).

\bibitem{Schafer:2007pr}
T.~Sch\"afer,
``The Shear Viscosity to Entropy Density Ratio of Trapped Fermions in the
Unitarity Limit,''
Phys.\ Rev.\  A {\bf 76}, 063618 (2007)
[arXiv:cond-mat/0701251].

\bibitem{Cao:2010wa}
C.~Cao, E.~Elliott, J.~Joseph, H.~Wu, J.~Petricka, 
T. Sch{\"a}fer, J.~E.~Thomas,
``Observation of Universal Temperature Scaling in the Quantum 
Viscosity of a Unitary Fermi Gas''
Science {\bf 331}, 58 (2011).
[arXiv:1007.2625 [cond-mat.quant-gas]].

\bibitem{Schafer:2009dj}
T.~Sch\"afer and D.~Teaney,
``Nearly Perfect Fluidity: From Cold Atomic Gases to Hot Quark Gluon
Plasmas,''
Rept.\ Prog.\ Phys.\  {\bf 72}, 126001 (2009)
[arXiv:0904.3107 [hep-ph]].

\bibitem{Adams:2012th} 
A.~Adams, L.~D.~Carr, T.~Sch\"afer, P.~Steinberg and J.~E.~Thomas,
``Strongly Correlated Quantum Fluids: Ultracold Quantum Gases, Quantum 
Chromodynamic Plasmas, and Holographic Duality,''
arXiv:1205.5180 [hep-th].

\bibitem{Policastro:2001yc}
G.~Policastro, D.~T.~Son and A.~O.~Starinets,
``The shear viscosity of strongly coupled N = 4 supersymmetric Yang-Mills
plasma,''
Phys.\ Rev.\ Lett.\  {\bf 87}, 081601 (2001)
[arXiv:hep-th/0104066].

\bibitem{Kovtun:2004de}
P.~Kovtun, D.~T.~Son and A.~O.~Starinets,
``Viscosity in strongly interacting quantum field theories from black hole
physics,''
Phys.\ Rev.\ Lett.\  {\bf 94}, 111601 (2005)
[arXiv:hep-th/0405231].

\bibitem{Landau:smII}
L.~D.~Landau, E.~M.~Lifshitz,
``Statistical Mechanics, Part II'', 
Course of Theoretical Physics, Vol.IX, 
Pergamon Press (1981).

\bibitem{Ma:1976}
S.-K.~Ma,
``Modern Theory Of Critical Phenomena,''
W.~A.~Benjamin (1976).

\bibitem{Onuki:2002}
A.~Onuki, 
``Phase Transition Dynamics,''
Cambridge University Press (2002).

\bibitem{Martin:1973zz} 
P.~C.~Martin, E.~D.~Siggia and H.~A.~Rose,
``Statistical Dynamics of Classical Systems,''
Phys.\ Rev.\ A {\bf 8}, 423 (1973).

\bibitem{DeDominicis:1977fw} 
C.~De Dominicis and L.~Peliti,
``Field Theory Renormalization and Critical Dynamics Above $T_c$: 
Helium, Antiferromagnets and Liquid Gas Systems,''
Phys.\ Rev.\ B {\bf 18}, 353 (1978).

\bibitem{Khalatnikov:1983ak} 
I.~M.~Khalatnikov, V.~V.~Lebedev and A.~I.~Sukhorukov,
``Diagram Technique For Calculating Long Wave Fluctuation Effects,''
Phys.\ Lett.\ A {\bf 94}, 271 (1983).

\bibitem{Kovtun:2011np} 
P.~Kovtun, G.~D.~Moore and P.~Romatschke,
``The stickiness of sound: An absolute lower limit on viscosity and 
the breakdown of second order relativistic hydrodynamics,''
Phys.\ Rev.\ D {\bf 84}, 025006 (2011)
[arXiv:1104.1586 [hep-ph]].

\bibitem{PeraltaRamos:2011es} 
J.~Peralta-Ramos and E.~Calzetta,
``Shear viscosity from thermal fluctuations in relativistic conformal 
fluid dynamics,''
JHEP {\bf 1202}, 085 (2012)
[arXiv:1109.3833 [hep-ph]].

\bibitem{Kovtun:2012rj} 
P.~Kovtun,
``Lectures on hydrodynamic fluctuations in relativistic theories,''
arXiv:1205.5040 [hep-th].

\bibitem{Torrieri:2011ne} 
G.~Torrieri,
``Viscosity of An Ideal Relativistic Quantum Fluid: A Perturbative study,''
Phys.\ Rev.\ D {\bf 85}, 065006 (2012)
[arXiv:1112.4086 [hep-th]].

\bibitem{Son:2005rv} 
D.~T.~Son and M.~Wingate,
``General coordinate invariance and conformal invariance in nonrelativistic 
physics: Unitary Fermi gas,''
Annals Phys.\  {\bf 321}, 197 (2006)
[cond-mat/0509786].

\bibitem{Son:2005tj}
D.~T.~Son,
``Vanishing bulk viscosities and conformal invariance of unitary Fermi gas,''
Phys.\ Rev.\ Lett.\  {\bf 98}, 020604 (2007)
[arXiv:cond-mat/0511721].

\bibitem{Chao:2010tk}
J.~Chao, M.~Braby, T.~Sch\"afer,
``Viscosity spectral functions of the dilute Fermi gas in kinetic theory,'' 
New J.\ Phys.\  {\bf 13}, 035014 (2011)
[arXiv:1012.0219 [cond-mat.quant-gas]].

\bibitem{Baier:2007ix} 
R.~Baier, P.~Romatschke, D.~T.~Son, A.~O.~Starinets and M.~A.~Stephanov,
``Relativistic viscous hydrodynamics, conformal invariance, and holography,''
JHEP {\bf 0804}, 100 (2008)
[arXiv:0712.2451 [hep-th]].

\bibitem{Braby:2010ec} 
M.~Braby, J.~Chao and T.~Sch\"afer,
``Thermal Conductivity and Sound Attenuation in Dilute Atomic Fermi Gases,''
Phys.\ Rev.\ A {\bf 82}, 033619 (2010)
[arXiv:1003.2601 [cond-mat.quant-gas]].

\bibitem{Bruun:2007}
G.~M.~Bruun, H.~Smith,
``Frequency and damping of the Scissors Mode of a Fermi gas,''
Phys. Rev. A {\bf 76}, 045602 (2007) 
[arXiv:0709.1617].

\bibitem{Massignan:2004}
P.~Massignan, G.~M.~Bruun, H.~Smith,
``Viscous relaxation and collective oscillations in a trapped Fermi 
gas near the unitarity limit,''
Phys.\ Rev.\ A {\bf 71}, 033607 (2005) 
[cond-mat/0409660].

\bibitem{Ku:2012}
M.~J.~H.~Ku, A.~T.~Sommer, L.~W.~Cheuk, and M.~W.~Zwierlein,
``Revealing the Superfluid Lambda Transition in the Universal
Thermodynamics of a Unitary Fermi Gas,''
Science 335, 563 (2012)
[arXiv:1110.3309 [cond-mat.quant-gas]].

\bibitem{Kinast:2005}
J.~Kinast, A.~Turlapov, and J.~E.~Thomas,
``Damping of a Unitary Fermi Gas,''
Phys.\ Rev.\ Lett.\  {94}, 170404 (2005)
[cond-mat/0502507].

\bibitem{Schaefer:2009px}
T.~Sch\"afer and C.~Chafin,
``Scaling Flows and Dissipation in the Dilute Fermi Gas at Unitarity,''
in: Springer Lecture Notes in Physics ``BEC-BCS Crossover and 
the Unitary Fermi gas,'' Wilhelm Zwerger (editor)
[arXiv:0912.4236 [cond-mat.quant-gas]].

\bibitem{Rupak:2007vp} 
G.~Rupak and T.~Sch\"afer,
``Shear viscosity of a superfluid Fermi gas in the unitarity limit,''
Phys.\ Rev.\ A {\bf 76}, 053607 (2007)
[arXiv:0707.1520 [cond-mat.other]].

\bibitem{Manuel:2011ed} 
C.~Manuel and L.~Tolos,
``Shear viscosity due to phonons in superfluid neutron stars,''
Phys.\ Rev.\ D {\bf 84}, 123007 (2011)
[arXiv:1110.0669 [astro-ph.SR]].

\bibitem{Enss:2010qh}
T.~Enss, R.~Haussmann, W.~Zwerger,
``Viscosity and scale invariance in the unitary Fermi gas,''
Annals Phys.\  {\bf 326}, 770-796 (2011).
[arXiv:1008.0007 [cond-mat.quant-gas]].

\bibitem{Wlazlowski:2012jb} 
G.~Wlazlowski, P.~Magierski and J.~E.~Drut,
``Shear Viscosity of a Unitary Fermi Gas,''
Phys.\ Rev.\ Lett.\  {\bf 109}, 020406 (2012)
[arXiv:1204.0270 [cond-mat.quant-gas]].

\bibitem{Ernst:1971}
M.~H.~Ernst, E.~H.~Hauge, J.~M.~J.~van Leeuwen 
``Asymptotic Time Behavior of Correlation Functions. I. Kinetic Terms,''
Phys.\ Rev.\ A {\bf 4}, 2055 (1971).

\bibitem{Forster:1977}
D.~Forster, D.~R.~Nelson, M.~J.~Stephen, 
``Large-distance and long-time properties of a randomly stirred fluid,''
Phys.\ Rev.\ A {\bf 16}, 732 (1977).

\bibitem{Khalatnikov:1984}
I.~M.~Khalatnikov, V.~V.~Lebedev, A.~I.~Sukhorukov,
``Fluctuation effects in two-dimensional hydrodynamic systems,''
Physica A  {\bf 126} 135 (1984).

\bibitem{Alder:1967}
B.~J.~Alder and T.~E.~Wainwright, 
``Velocity autocorrelations for hard spheres,''
Phys.\ Rev.\ Lett.\ {\bf 18}, 988 (1967).

\bibitem{Kadanoff:1989}
L.~P.~Kadanoff, G.~R.~McNamara, and G.~Zanetti, 
``From automata to fluid flow: Comparisons of simulation and theory,''
Phys.\ Rev.\ A {\bf 40}, 4527 (1989). 

\bibitem{Schafer:2011my} 
T.~Sch\"afer,
``Shear viscosity and damping of collective modes in a two-dimensional 
Fermi gas,''
Phys. Rev. A {\bf 85}, 033623 (2012)
[arXiv:1111.7242 [cond-mat.quant-gas]].

\bibitem{Bruun:2011}
G.~M.~Bruun, 
``Shear viscosity and spin-diffusion coefficient of a two-dimensional 
Fermi gas,''
Phys. Rev. A {\bf 85}, 013636 (2012) 
[arXiv:1112.2395 [cond-mat.quant-gas]].

\bibitem{Randeria:1990}
M.~Randeria, J.-M.~Duan, and L.-Y.~Shieh, 
``Superconductivity in a two-dimensional Fermi gas: Evolution from Cooper 
pairing to Bose condensation,'' 
Phys.\ Rev.\ B {\bf 41}, 327 (1990). 

\bibitem{Vogt:2011}
E.~Vogt, M.~Feld, B.~Fr\"ohlich, D.~Pertot, M.~Koschorreck, M.~K\"ohl,
``Scale invariance and viscosity of a two-dimensional Fermi gas,''
Phys.\ Rev.\ Lett.\ {\bf 108}, 070404 (2012)
[arXiv:1111.1173 [cond-mat.quant-gas]].

\bibitem{Cao:2011fh}
C.~Cao, E.~Elliott, H.~Wu and J.~E.~Thomas,
``Searching for Perfect Fluids: Quantum Viscosity in a Universal Fermi Gas,''
New J.\ Phys.\  {\bf 13} (2011) 075007
[arXiv:1105.2496 [cond-mat.quant-gas]].

\bibitem{Kinast:2004b}
J.~Kinast, A.~Turlapov and J.~E.~Thomas,
``Breakdown of Hydrodynamics in the Radial Breathing Mode of a
Strongly-Interacting Fermi Gas,''
Phys.\ Rev.\ A {70}, 051401(R) (2004)
[cond-mat/0408634].

\bibitem{Bartenstein:2004}
M.~Bartenstein, A.~Altmeyer, S.~Riedl, S.~Jochim, C.~Chin,
J.~Hecker Denschlag, and R.~Grimm,
``Collective Excitiations of a Degenerate Gas at the BEC-BCS Crossover,''
Phys.\ Rev.\ Lett.\ {92}, 203201 (2004)
[cond-mat/0412712].

\bibitem{Altmeyer:2006}
A.~Altmeyer, S.~Riedl, C.~Kohstall, M.~Wright, R.~Geursen, M.~Bartenstein,
C.~Chin, J.~Hecker Denschlag, and R.~Grimm,
``Precision Measurements of Collective Oscillations in the BEC-BCS Crossover,''
Phys.\ Rev.\ Lett.\ {98}, 040401 (2007)
[cond-mat/0609390].

\bibitem{Luo:2009}
L.~Luo and J.~E. Thomas,
``Thermodynamic measurements in a strongly interacting Fermi gas,''
J.\ Low Temp.\ Phys.\ {\bf 154}, 1 (2009)
[arXiv:0811.1159 [cond-mat.other]].

\bibitem{Mannarelli:2012su}
M.~Mannarelli, C.~Manuel and L.~Tolos,
``Shear viscosity in a superfluid cold Fermi gas at unitarity,''
arXiv:1201.4006 [cond-mat.quant-gas].

\bibitem{Balboa:2011}
F.~B.~Balboa, J.~B.~Bell, R.~Delgado-Buscalioni, A.~Donev, T.~G.~Fai, 
B.~E.~Griffith, C.~S.~Peskin,
``Staggered Schemes for Fluctuating Hydrodynamics,''
arXiv:1108.5188 [physics.flu-dyn].

\bibitem{Bruun:2007b}
G. M. Bruun and H. Smith 
Shear viscosity and damping for a Fermi gas in the unitarity limit
Phys. Rev. A 75, 043612 (2007)
[arXiv:cond-mat/0612460].

\bibitem{Guo:2010dc} 
H.~Guo, D.~Wulin, C.~-C.~Chien and K.~Levin,
``Microscopic Approach to Shear Viscosities in Superfluid Gases: 
From BCS to BEC,''
Phys.\ Rev.\ Lett.\  {\bf 107}, 020403 (2011)
[arXiv:1008.0423 [cond-mat.quant-gas]].

\bibitem{Guo:2010xu} 
H.~Guo, D.~Wulin, C.-C.~Chien and K.~Levin,
``Perfect Fluids and Bad Metals: Transport Analogies Between Ultracold 
Fermi Gases and High $T_c$ Superconductors,''
New J.\ Phys.\  {\bf 13}, 075011 (2011)
[arXiv:1009.4678 [cond-mat.supr-con]].

\bibitem{Romatschke:2012}
P.~Romatschke, R.~E.~Young, 
Comment on ``Hydrodynamic fluctuations and the minimum shear 
viscosity of the dilute Fermi gas at unitarity'',
arXiv:1209.1604 [cond-mat.quant-gas].

\bibitem{Burgess:1998ku} 
C.~P.~Burgess,
``Goldstone and pseudo-Goldstone bosons in nuclear, particle and 
condensed matter physics,''
Phys.\ Rept.\  {\bf 330}, 193 (2000)
[hep-th/9808176].

\bibitem{Kaplan:1998tg} 
D.~B.~Kaplan, M.~J.~Savage and M.~B.~Wise,
``A New expansion for nucleon-nucleon interactions,''
Phys.\ Lett.\ B {\bf 424}, 390 (1998)
[nucl-th/9801034].

\bibitem{Fox:1970}
R.~F.~Fox and G.~E.~Uhlenbeck,
``Contributions to Non-Equilibrium Thermodynamics. I. Theory of
Hydrodynamical Fluctuations,''
Phys.\ Fluids {\bf 13} 1893 (1970).

\end{thebibliography}
\end{document}